\documentclass[aps,pra,twocolumn,superscriptaddress,showpacs]{revtex4}
\usepackage{amsmath}
\usepackage{epsfig}

\newcommand{\beq}{\begin{eqnarray}}
\newcommand{\eeq}{\end{eqnarray}}

\begin{document}

\title{Detecting Bose-Einstein condensation of exciton-polaritons via
electron transport }
\author{Yueh-Nan Chen}
\email{yuehnan@mail.ncku.edu.tw}
\affiliation{Department of Physics and National Center for Theoretical Sciences, National
Cheng-Kung University, Tainan 701, Taiwan}
\author{Neill Lambert}
\affiliation{Advanced Science Institute, The Institute of Physical and Chemical Research
(RIKEN), Saitama 351-0198, Japan}
\author{Franco Nori}
\affiliation{Advanced Science Institute, The Institute of Physical and Chemical Research
(RIKEN), Saitama 351-0198, Japan}
\affiliation{Physics Department and Center for Theoretical Physics, The University of
Michigan, Ann Arbor, M1 48109-1040, USA}
\date{\today}

\begin{abstract}
We examine the Bose-Einstein condensation of exciton-polaritons in a
semiconductor microcavity via an electrical current. We propose that
by embedding a quantum dot \textit{p-i-n} junction inside the
cavity, the tunneling current through the device can reveal features
of condensation due to a one-to-one correspondence of the photons to
the condensate polaritons. Such a device can also be used to observe
the phase interference of the order parameters from two condensates.
\end{abstract}

\maketitle





\section{Introduction}The essence of Bose-Einstein condensation
(BEC) is the macroscopic occupation of a single-particle
state$^{1,2}$. The achievement of BEC in dilute atomic gases has
enabled the study of the long-range spatial coherence in a
well-controlled environment$^{2}$. In contrast to the extremely low
temperatures needed for dilute atom gases, excitons in
semiconductors have long been considered a candidate for BEC at
temperatures of a few Kelvin, due to their light effective
mass$^{3}$. In the past few decades, numerous studies have shown
evidence$^{4}$ for the existence of excitonic BEC. A recent
promising realization for such a BEC is within a two-dimensional
quantum well in a microcavity, i.e., \emph{a condensate of
polaritons}$^{5}$, which are half-light, half-matter bosonic
quasi-particles. Fascinating features of condensate polaritons, such
as
phase interference$^{6}$, quantized vortices$^{7}$, Bogoliubov excitations$%
^{8}$, and collective fluid dynamics$^{9}$, have been successfully observed
in experiments.

In a context related to the study of semiconductor microcavities, an exciton
in a quantum dot (QD) embedded inside a microcavity can be used to study the
phenomena of cavity quantum electrodynamics$^{10}$. With the advances of
fabrication and measuring technologies, strong couplings between the QD
excitons and cavity photons have been observed both in a semiconductor
microcavity$^{11}$ and in a photonic crystal nanocavity$^{12}$. Another
unique feature of artificial atoms, such as QDs, is that they can be
connected to electronic reservoirs. For example, it is now possible to embed
QDs inside a \textit{p-i-n} structure$^{13}$, such that electrons and holes
can be injected separately from opposite sides. This allows one to examine
the exciton dynamics in a QD via electrical currents$^{14}$.

Motivated by these recent developments, 
we propose a method to detect the BEC of polaritons via an electrical
current by embedding a QD \textit{p-i-n}\textrm{\ }junction inside a
microcavity, where the condensation of polaritons takes place. This is in
principle feasible since the excitation energy of the QD exciton (two-level
spacing) is comparable to that of the cavity photons. Once the condensation
of polaritons occurs, the one-to-one correspondence between the polariton
and its half-light part (photon) ensures that the photons also condense to
their ground state. In this case, the transport current through the dot
should ``feel'' the condensation. We will show that the contribution to the
\emph{coherent} transport of the current increases with the condensate
fraction. Furthermore, if the QD is coupled to two condensates, the
current-noise can reveal the phase interference between them.
\begin{figure}[h]
\includegraphics[width=\columnwidth]{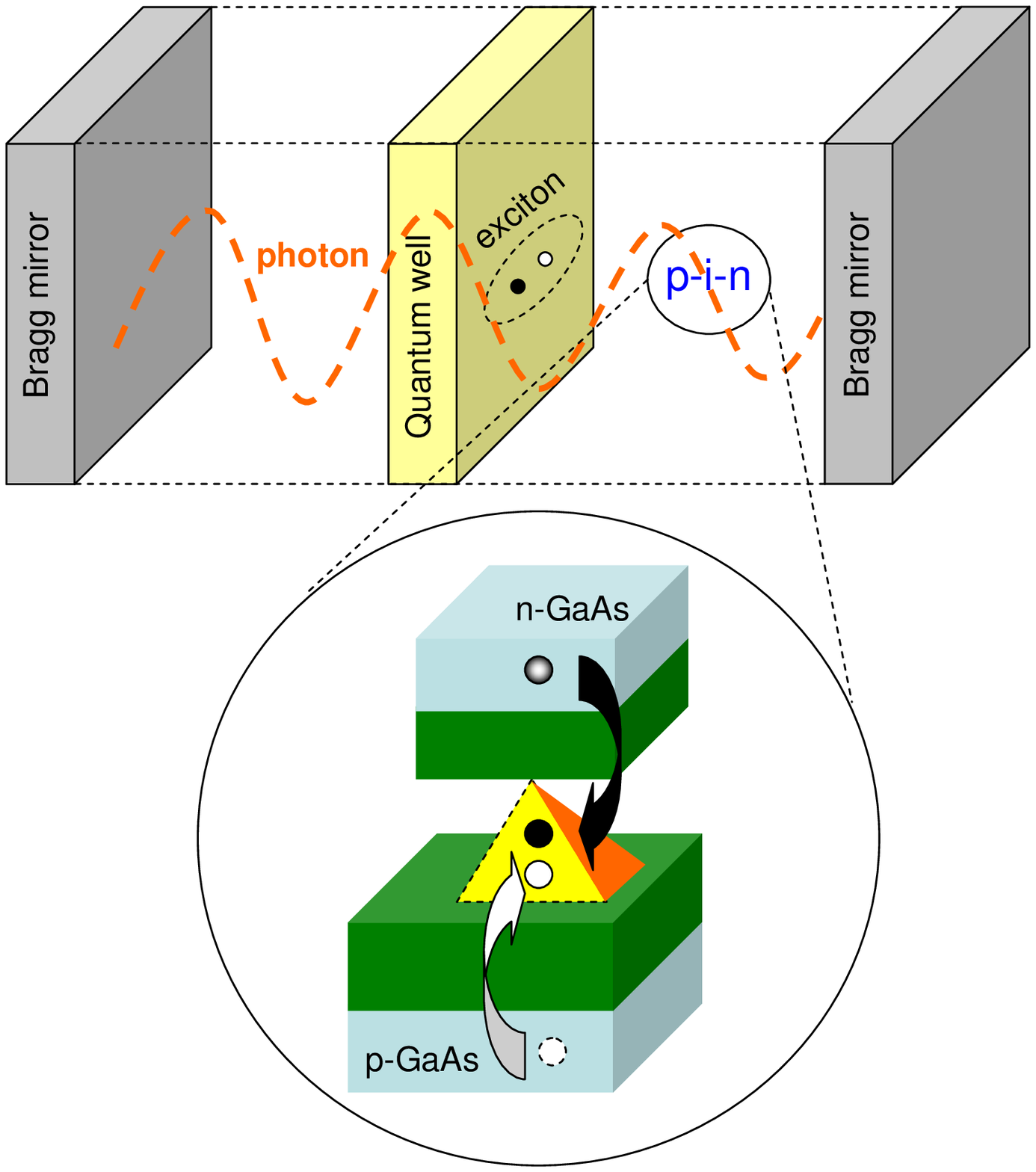}
\caption{{}(Color online) Schematic diagram of the system: a semiconductor
quantum well is placed between two Bragg mirrors. A quantum dot \textit{p-i-n%
} junction is embedded between the well and the mirror to detect the photon
part of the polaritons. For simplicity, the substrate of the QDs is not
shown. }
\end{figure}

\section{Quantum dot p-i-n junction in a microcavity}Consider now
a QD \textit{p-i-n} junction embedded inside a semiconductor
microcavity, where the quantum well excitons and cavity photons
condense to their ground state as shown in Fig. 1. When this
condensation occurs, a great number of
polaritons, $\widehat{b}_{\mathbf{k}}$, will occupy the zero-momentum state $%
\mathbf{k}_{0}$. The canonical transformation$^{1,2}$

\begin{equation}
\widehat{b}_{\mathbf{k}}=\sqrt{N^{\prime }}e^{i\phi }\delta _{\mathbf{k},%
\mathbf{k}_{0}}+\widehat{\alpha }_{\mathbf{k}}
\end{equation}%
is commonly used to describe $N^{\prime }$ condensed particles and
non-condensate particles with operator $\widehat{\alpha }_{\mathbf{k}}$. The
polariton operator $\widehat{b}_{\mathbf{k}}$ is composed of the exciton
operator, $\widehat{c}_{\mathbf{k}}$, and photon operator, $\widehat{a}_{%
\mathbf{k}}$,

\begin{equation}
\widehat{b}_{\mathbf{k}}=u_{\mathbf{k}}\widehat{c}_{\mathbf{k}}+v_{\mathbf{k}%
}\widehat{a}_{\mathbf{k}},
\end{equation}%
where $u_{\mathbf{k}}$ and $v_{\mathbf{k}}$ are coefficients easily
obtained from the diagonalization of exciton-photon
interaction$^{15}$. From Eq.~(2), we can see that there is a
one-to-one correspondence of the polariton operator to the photon
one. Therefore, the canonical transformation in Eq.~(1) can also be applied to the photon operator: $$\widehat{a}_{\mathbf{k}}=%
\sqrt{N}e^{i\varphi }\delta
_{\mathbf{k},\mathbf{k}_{0}}+\widehat{\beta }_{\mathbf{k}\neq
\mathbf{k}_{0}},$$ where $\widehat{\beta }_{\mathbf{k}}$ represents
the photons not in the zero momentum state. The photon condensate
fraction $N$ is related to $N^{\prime }$ via the particular choice
of the diagonalization in Eq. (2).

In this case, the exciton-photon interaction in the QD \textit{p-i-n}
junction, $H_{\mathrm{ex-ph}}$, can now be written as%
\begin{equation}
H_{\mathrm{ex-ph}}=T_{0}e^{i\varphi }\left| \uparrow \right\rangle
\left\langle \downarrow \right| +\sum_{\mathbf{k}\neq \mathbf{k}_{0}}D_{%
\mathbf{k}}\left| \uparrow \right\rangle \left\langle \downarrow \right|
\widehat{\beta }_{\mathbf{k}}+\text{H.c.},
\end{equation}%
where $T_{0}=\sqrt{N}D_{\mathbf{k}_{0}}$, with $D_{\mathbf{k}}$ being the
coupling strength between the dot exciton and the cavity photon. The index $%
\mathbf{k}_{0}$ represents the condensate ground state.

We have essentially assumed a mean-field interaction, so that the
field of the condensate mode is just represented by a $c$-number:
there is no backaction from the QD to the cavity.  From the theory
of transport through QDs, the first term in Eq. (3) represents
\emph{coherent}
tunneling$^{16}$, while the second term describes \emph{incoherent} tunneling%
$^{14}$. Here, we have introduced the three dot states: $\left|
0\right\rangle =\left| 0,h\right\rangle $, $\left| \uparrow \right\rangle
=\left| e,h\right\rangle $, and $\left| \downarrow \right\rangle =\left|
0,0\right\rangle $, where $\left| 0,h\right\rangle $ means that there is one
hole in the QD,\ $\left| e,h\right\rangle $ is the exciton state, and $%
\left| 0,0\right\rangle $ represents the ground state with no hole and no
electron in the QD$^{14}$. The Hamiltonian describing the tunneling to the
electron and hole reservoirs can thus be written as%
\begin{equation}
H_{T}=\sum_{\mathbf{q}}(V_{\mathbf{q}}\widehat{d}_{e,\mathbf{q}}^{\dagger
}\left| 0\right\rangle \left\langle \uparrow \right| +W_{\mathbf{q}}\widehat{%
d}_{h,\mathbf{q}}^{\dagger }\left| 0\right\rangle \left\langle \downarrow
\right| +\text{H.c.}),
\end{equation}
where $\widehat{d}_{e,\mathbf{q}}$ and $\widehat{d}_{h,\mathbf{q}}$ are the
electron operators in the electron and hole reservoirs, respectively. Here, $%
V_{\mathbf{q}}$ and $W_{\mathbf{q}}$ couple the channel with momentum $%
\mathbf{q}$ of the electron and the hole reservoirs.

One can now write the equation of motion for the reduced density operator:

\begin{eqnarray}
\frac{d}{dt}\rho (t) &=&-i[H_{\mathrm{coh}}(t),\rho (t)]-\mathrm{Tr}_{%
\mathrm{res}}\int_{0}^{t}dt^{\prime }[H_{\mathrm{incoh}}(t)  \notag \\
&&+H_{T}(t),[H_{\mathrm{incoh}}(t^{\prime })+H_{T}(t^{\prime }),\widetilde{%
\Xi }(t^{\prime })]],
\end{eqnarray}%
where $\widetilde{\Xi }(t^{\prime })$ is the total density operator, and $H_{%
\mathrm{coh}}$ ($H_{\mathrm{incoh}}$) represents the coherent (incoherent)
tunneling in Eq. (3). Note that the trace, $\mathrm{Tr}$, in Eq.~(5) is
taken with respect to both the non-condensate photons and the electronic
reservoirs.

\section{Tunneling current}If the couplings to the non-condensate
photons and to the electron/hole reservoirs are weak, then it is
reasonable to assume that the standard Born-Markov approximation
with respect to these couplings is valid. In this case, one can
derive a master equation from the exact time-evolution of the system
and obtain the tunnel current through the
hole-side barrier$^{14}$: $I(t)\equiv -e\Gamma _{R}\left\langle \widehat{n}%
_{\downarrow }\right\rangle _{t}$, where $\widehat{n}_{\downarrow }=\left|
\downarrow \right\rangle \left\langle \downarrow \right| $ and $\Gamma _{R}$
is the tunneling rate from the hole side reservoir.

In the steady state limit ($t\rightarrow \infty $), the analytical
expression for the tunneling current $I$ is given by
\begin{equation}
I(t\rightarrow \infty )=\frac{2\Gamma _{R}T_{0}^{2}+2\gamma \varepsilon ^{2}%
}{[\varepsilon ^{2}+T_{0}^{2}(2+\Gamma _{R}/\Gamma _{L})]+\gamma \varepsilon
^{2}(1/\Gamma _{R}+1/\Gamma _{L})},
\end{equation}%
where $\varepsilon ^{2}=E_{0}^{2}+\Gamma _{R}^{2}$. Here, $E_{0}$ is the
quantum dot exciton bandgap, $\Gamma _{L}$ is the tunneling rate from the
electron-side reservoir, and $\gamma $ is the incoherent decay rate due to
the non-condensate photons. Note that, for convenience, we have set the
electron charge $e=1$ and Planck constant $\hbar =1$.

Examining Eq. (6) we note that, when the condensation number $N$($\propto
T_{0}^{2}$) becomes relatively large, the steady-state current $%
I(t\rightarrow \infty )$ saturates to the value:

\begin{equation}
I(t\rightarrow \infty )\underset{N\rightarrow \infty }{\longrightarrow }%
\frac{\Gamma _{R}}{1+\frac{\Gamma _{R}}{2\Gamma _{L}}},
\end{equation}%
depending only on the values of the tunneling rates $\Gamma _{L}$ and $%
\Gamma _{R}$. In the opposite limit of no condensation, Eq. (6) is reduced
to the result of incoherent case$^{17}$:
\begin{equation}
I(t\rightarrow \infty )\underset{T_{0}\rightarrow 0}{\longrightarrow }\left(\frac{%
1}{\Gamma _{R}}+\frac{1}{\Gamma _{L}}+\frac{1}{\gamma }\right)^{-1}.
\end{equation}%
The curve in the inset of Fig. 2 shows that the current $I$\ increases when
increasing the occupation number $N$. Such a phenomenon may be observed by
increasing the power of the laser excitation, as has been performed in
experiments$^{5}$. Note that in the inset of Fig. 2 and the following
figures, we have set the exciton bandgap $E_{0}=1.4$ $\mathrm{eV}$ \ and the
tunneling rates: $\Gamma _{R}=10\Gamma _{L}=0.1$ $\mathrm{meV}$.
\begin{figure}[h]
\includegraphics[width=\columnwidth]{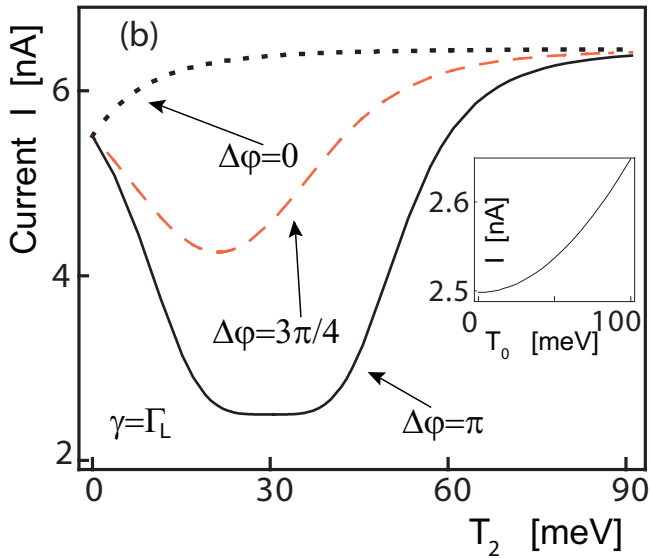}
\caption{{}(Color online) Under the influence of two independent
condensates, the dotted, red-dashed, and continuous black curves represent
the current through the QD \textit{p-i-n} junction for different values of
phase difference: $\protect\varphi _{1}-\protect\varphi _{2}=0$, $3\protect%
\pi /4$, and $\protect\pi $, respectively. In plotting the figure,
the values of $T_{1}$ $(=30 \mathrm{meV})$ and incoherent rate
$\protect\gamma$ $(\gamma=\Gamma _{L})$ are kept fixed. The inset
shows the current increases when increasing the condensation number
$N$ ($\propto T_{0}^{2}$) for the case of only a single condensate.
$T_{0}$ is the coupling strength between the dot exciton and the
cavity photon.\ }
\end{figure}

\section{Interference between two condensates}Another important
effect that can be examined is the interference between two
condensates, which has been observed and verified in dilute atomic
gases$^{18}$. Consider now an additional quantum well in the
microcavity, so that the excitons in this well also form a
condensate with the photons. The interactions experienced
by the \textit{p-i-n} junction experiences can be described by%
\begin{equation}
H_{\mathrm{2ex-ph}}=\sum_{j=1,2}\{T_{j}e^{i\varphi _{j}}\left| \uparrow
\right\rangle \left\langle \downarrow \right| +\sum_{\mathbf{k}\neq \mathbf{k%
}_{0}}D_{j,\mathbf{k}}\left| \uparrow \right\rangle \left\langle \downarrow
\right| \widehat{\beta }_{j,\mathbf{k}}\}+\text{H.c.},
\end{equation}%
where the two phases $\varphi _{1}$ and $\varphi _{2}$ come from the $U(1)$
symmetry-breaking of the two condensates. Assuming that the exciton-photon
couplings of the two wells are identical, the \emph{coherent} parts, $T_{j}=%
\sqrt{N_{j}}D_{\mathbf{k}_{0}}$, contain the information of the excitation
numbers $N_{1}$ and $N_{2}$. The resultant steady-state current is similar
to Eq. (6), besides the following replacement:%
\begin{equation}
T_{0}^{2}\rightarrow D_{\mathbf{k}_{0}}^{2}[N_{1}+2\sqrt{N_{1}N_{2}}\cos
(\varphi _{1}-\varphi _{2})+N_{2}].
\end{equation}%
For a fixed $N_{1}$, the dotted, red-dashed, and black curves in Fig. 1(b)
represent the steady-state currents as functions of $N_{2}$, for the phase
differences $\varphi _{1}-\varphi _{2}=0$, $3\pi /4$, and $\pi $,
respectively. As seen in Fig. 2, the dips in the currents reveal the effect
of destructive interference when $\varphi _{1}-\varphi _{2}$ approaches $\pi
$.

We also suggest that the \textit{p-i-n} junction can be embedded inside an
array of polariton condensates connected by weak periodic potential barriers$%
^{6}$, where the in-phase (`zero-state') and anti-phase (`$\pi $-state')
have been created. In this case, Eqs. (6) and (10) can also be used to
distinguish the zero-state and $\pi $-state.
\begin{figure}[h]
\includegraphics[width=7.5cm]{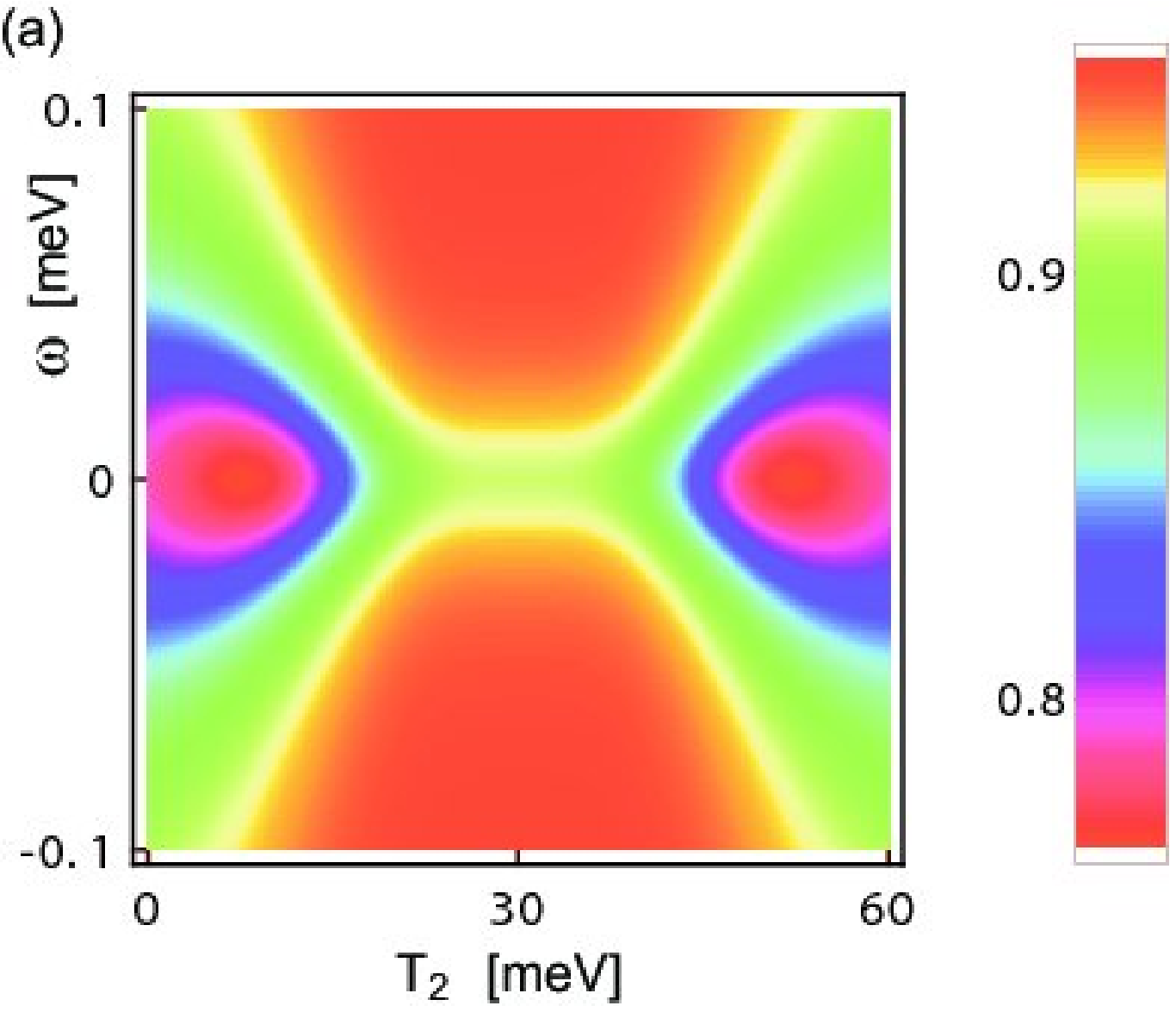} %
\includegraphics[width=7.5cm]{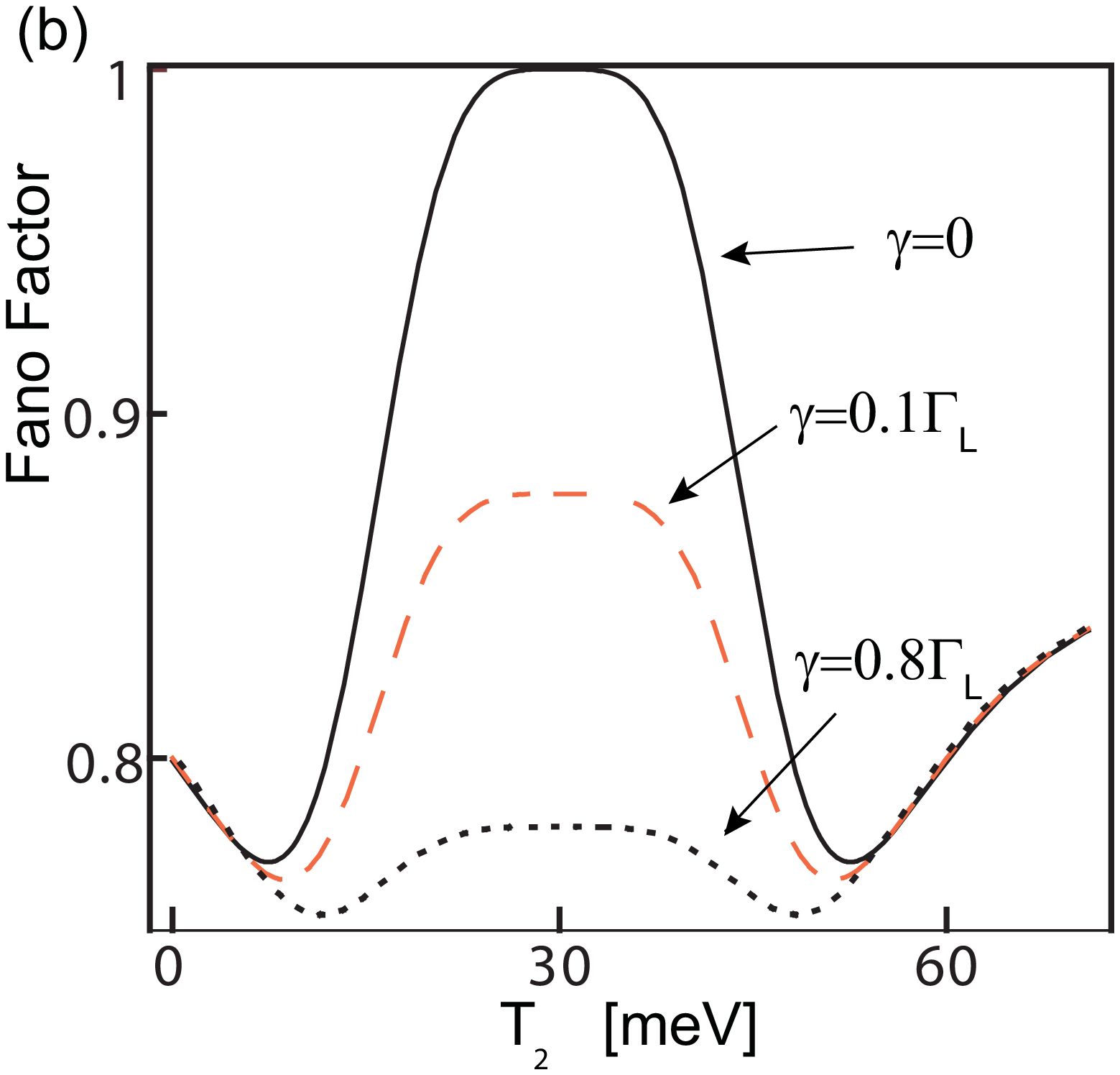}
\caption{{}(Color online) (a) Shot-noise spectra of a QD \textit{p-i-n}
junction as functions of both $\protect\omega $ and the coupling $T_{2}$ to
the 2$^{nd}$ condensate. Here, the excitons are coupled to two condensates.
Similar to Fig. 1(b), one of the condensate numbers is kept fixed, i.e. $%
T_{1}=30$ $\mathrm{meV}$. (b) By fixing the phase difference $\protect%
\varphi _{1}-\protect\varphi _{2}=\protect\pi $, the continuous black, red
dashed, and dotted curves represent the Fano factor (zero-frequency noise)
for $\protect\gamma =0$, $0.1\Gamma _{L}$, and $0.8\Gamma _{L}$,
respectively. The Fano factor is defined here as $S_{I_{R}}(\protect\omega %
\rightarrow 0)/2eI$.}
\end{figure}

\section{Shot-noise measurements}Recently, interest in
measurements of shot-noise in quantum transport has grown owing to
the possibility of extracting valuable information not available in
conventional dc transport experiments$^{19}$. Therefore, in addition
to the current, we now proceed to calculate the noise spectrum.

In a quantum conductor out of equilibrium, electronic current-noise
originates from the dynamical fluctuations of the current away from
its average. To study correlations between carriers, we relate the
exciton dynamics with the hole reservoir operators by introducing
the degree of freedom $n$ as the number of holes that have tunneled
through the barrier connected to the reservoir of holes and write

\begin{eqnarray}
\overset{\cdot }{n}_{0}^{(n)}(t) &=&-\Gamma _{L}n_{0}^{(n)}(t)+\Gamma
_{R}n_{\downarrow }^{(n-1)}(t), \\
\overset{\cdot }{n}_{\uparrow }^{(n)}(t) &=&\Gamma
_{L}n_{0}^{(n)}(t)+iT[p_{\uparrow ,\downarrow }^{(n)}(t)-p_{\downarrow
,\uparrow }^{(n)}(t)]-\gamma n_{\uparrow }^{(n)}(t),  \notag \\
\overset{\cdot }{n}_{\downarrow }^{(n)}(t) &=&-\Gamma
_{R}n_{0}^{(n)}(t)-iT[p_{\uparrow ,\downarrow }^{(n)}(t)-p_{\downarrow
,\uparrow }^{(n)}(t)]+\gamma n_{\uparrow }^{(n)}(t),  \notag
\end{eqnarray}%
where $n_{j}^{(n)}(t),$ $j=$ $0$, $\uparrow ,$ $\downarrow $, represent the
time-dependent occupation probabilities for the diagonal elements: $\left|
0\right\rangle \left\langle 0\right| $, $\left| \uparrow \right\rangle
\left\langle \uparrow \right| $, and $\left| \downarrow \right\rangle
\left\langle \downarrow \right| $, respectively. Here, $p_{\uparrow
,\downarrow }(t)$ and $p_{\downarrow ,\uparrow }(t)$ are the off-diagonal
matrix elements: $\left| \uparrow \right\rangle \left\langle \downarrow
\right| $ and $\left| \uparrow \right\rangle \left\langle \downarrow \right|
$. $T$ is the ``coherent'' interaction that the dot experiences. The
superscript `$n$' in $n_{j}^{(n)}(t)$ refers to the $n$ holes that have
tunneled the barrier connecting to the hole reservoir.

The Eqs. (11) allow us to calculate the particle current and the noise
spectrum $S_{I_{R}}(\omega )$ from $P_{n}(t)=n_{0}^{(n)}(t)+n_{\uparrow
}^{(n)}(t)+n_{\downarrow }^{(n)}(t)$ which gives the total probability of
finding $n$ electrons in the collector at time $t$. In particular, the noise
spectrum $S_{I_{R}}$ can be calculated via the MacDonald formula$^{20}$

\begin{equation}
S_{I_{R}}(\omega )=2\omega e^{2}\int_{0}^{\infty }dt\sin (\omega t)\frac{d}{%
dt}[\left\langle n^{2}(t)\right\rangle -(t\left\langle I\right\rangle )^{2}],
\end{equation}%
where $\frac{d}{dt}\left\langle n^{2}(t)\right\rangle =\sum_{n}n^{2}\overset{%
\cdot }{P_{n}}(t)$. From Eqs. (11) and (12), we obtain

\begin{equation}
S_{I_{R}}(\omega )=2eI\{1+\Gamma _{R}[\widetilde{n}_{\downarrow }(-i\omega )+%
\widetilde{n}_{\downarrow }(i\omega )]\},
\end{equation}%

where $\widetilde{n}_{\downarrow }(z)$ is the Laplace transformation of $%
n_{\downarrow }(t)$.

By fixing $T_{1}=30$ meV and $\varphi _{1}-\varphi _{2}=\pi $, an
interference effect can be observed in the noise spectrum as a function of $%
T_{2}$ and $\omega $, as shown in Fig. 3(a). The figure shows two
symmetric lobes around $T_{2}=T_{1}$, which represent the local
minima. To understand these features, we plot in Fig. 3(b) the
\textit{Fano factor} (i.e., the \textit{zero-frequency noise}
divided by the current) as a function of $T_{2}$ for different
values of the incoherent decay rate $\gamma $. One clearly finds
that the magnitude of the central peak decreases when increasing
$\gamma $. As the incoherent process dominates due to the
non-condensate photons overwhelming the coherent ones, therefore the
Fano factor reduces to the usual sub-Poissonian limit$^{21}$. In the
opposite limit ($\gamma \rightarrow 0$), the Fano factor approaches
unity, i.e. the Poissonian value, demonstrating that\ the revealing
feature of destructive interference is a peak in the Fano factor (at
$\omega =0$), coinciding with the dip in the steady-state current
observed in Fig. 1(b).

\section{Concluding remarks} An alternative way to detect the
condensation of polaritons via electron transport is by directly
embedding the quantum well between a \textit{p-n} junction. We
expect that in this case the increase of the steady-state current
with the condensation number $N$ might be observable.   However, the
features we observed in the current-noise spectrum may be invisible
since the assumption of three dot states is not valid in a quantum
well.

In summary, we have shown that a single QD \textit{p-i-n} junction
can serve as a mini-detector$^{22}$ inside the quantum structure,
such that the features of condensation and interference can be
readout via the electrical current and current-noise.

\acknowledgements

We would like to thank S. A. Gurvitz for helpful discussions. This
work is supported partially by the National Science Council of
Taiwan under the grant number 95-2112-M-006-031-MY3. FN acknowledges
partial support from the National Security Agency (NSA), Laboratory
for Physical Sciences (LPS), Army Research Office (ARO), National
Science Foundation (NSF) Grant No. EIA-0130383, and JSPS-RFBR
contract No. 06-02-91200.

\end{document}